\documentclass[12pt]{article}
\usepackage{graphicx}
\usepackage{dcolumn}
\usepackage{bm}
\usepackage{rotate}
\usepackage{epsfig}
\newcommand \be  {\begin{equation}}
\newcommand \bea {\begin{eqnarray} \nonumber }
\newcommand \ee  {\end{equation}}
\newcommand \eea {\end{eqnarray}}
\begin{document}
\title{Large dimension forecasting models and random singular value spectra}
\author{Jean-Philippe Bouchaud$^{1,2}$, Laurent Laloux$^1$ \\
M. Augusta Miceli$^{3,1}$, Marc Potters$^1$}
\maketitle
{\small{$ $\\
$^1$ Science \& Finance, Capital Fund Management, 6 Bd
Haussmann, 75009 Paris, France.\\
$^2$ Service de Physique de l'{\'E}tat Condens{\'e}
Orme des Merisiers -- CEA Saclay, 91191 Gif sur Yvette Cedex, France\\
$^3$ Universit\`a La Sapienza, Piazzale Aldo Moro, 00185 Roma, Italy}}



\begin{abstract}
We present a general method to detect and extract 
from a finite time sample statistically meaningful correlations between 
input and output variables of large dimensionality. Our central 
result is derived from the theory of free random matrices, and gives an explicit expression for the interval where
singular values are expected in the absence of any true correlations between the variables under study.
Our result can be seen as the natural generalization of the Mar\v{c}enko-Pastur distribution for the case of 
rectangular correlation matrices. We illustrate the interest of our method on a set of macroeconomic time series. 
\end{abstract}

\maketitle

\section{Introduction}

Finding correlations between observables is at the heart of scientific methodology. Once correlations between ``causes''
and ``effects'' are empirically established, one can start devising theoretical models to understand 
the mechanisms underlying such correlations, and use these models for prediction purposes. In many cases, 
the number of possible causes and of resulting effects are both large. For example, one can list an a priori large 
number of environmental factors possibly favoring the appearance 
of several symptoms or diseases, or of social/educational factors determining choices and tastes on different topics. A vivid example is provided by Amazon.com, where taste correlations between a huge number of 
different products (books, CDs, etc.) are sought for. In 
the context of gene expression networks, the number of input and output chemicals and proteins, described by their 
concentration, is very large. In an industrial setting, 
one can monitor a large number of characteristics of a device (engine, hardware, etc.) during the production phase and 
correlate these with the performances of the final product. In economics and finance, one aims at understanding the 
relation between a large number of possibly relevant factors, such as interest and exchange rates, industrial production, 
confidence index, etc. on, say, the evolution of inflation in different sectors of activity \cite{SW05}, or on 
the price of different stocks. Nowadays, the number of macroeconomic time series available to economists is huge 
(see below). This has lead Granger \cite{Granger} and others to suggest that ``large models'' should be at the forefront 
of the econometrics agenda. The theoretical study of high dimensional factor models is indeed actively pursued 
\cite{Gew,SW,Forni,Bai,Bai2,SW05}, in particular in relation with monetary policy \cite{Bernanke,SW05}. 

In the absence of information on the phenomenon under study, 
a brute force strategy would consist in listing a large number of possible explanatory variables and a large number of 
output variables, and systematically look for correlations between pairs, in the hope of finding some significant signal. 
In an econometric context, this is the point of view advocated long ago by Sims \cite{Sims}, who suggested to look at large 
Vector Autoregressive models, and let the system itself determine the number and the nature of the relevant variables. However, this procedure is rapidly affected by 
the ``dimensionality curse'', also called the problem of sunspot variables in the economics literature \cite{Woodford}. 
Since the number of observations is always limited, it can happen that two totally unrelated 
phenomenon (such as, for example, stock prices and sunspots) appear to be correlated over a certain time interval $T$. More
precisely, the correlation coefficient $\rho$, which would (presumably) be zero if very long time series could be studied, is in
fact of the order of $1/\sqrt{T}$ and can be accidentally large. When one tries to correlate systematically $N$ input
variables with $M$ output variables, the number of pairs is $NM$. In the absence of any true correlation between these
variables, the largest of these $NM$ empirical correlation coefficients will be, for Gaussian variables, of order 
$\rho_{\max} \sim \sqrt{2 \ln(NM)/T}$, which {\it grows} with $NM$. For example, $\rho_{\max} \approx 0.25$ for $N=M=25$ 
and $T=200$. If the input and output variables are non Gaussian and have fat-tails, this number can be even larger: 
if two strongly fluctuating random variable accidentally take large values simultaneously, this will contribute a lot 
to the empirical correlation even though $\rho$ should be zero for large $T$.

In this paper we want to discuss how recent results in Random Matrix Theory \cite{Verdu,Edelman} allow one to alleviate 
this dimensionality curse and give a precise procedure to extract significant correlations between $N$ input variables 
and $M$ output variables, when the number of independent observations is $T$. The idea is to compare the singular value 
spectrum of the empirical rectangular $M \times N$ correlation matrix with a benchmark, obtained by assuming no 
correlation at all between the variables. For $T \to \infty$ at $N,M$ fixed, all singular values should
be zero, but this will not be true if $T$ is finite.  
The singular value spectrum of this benchmark problem can in fact 
be computed exactly in the limit where $N,M,T \to \infty$, 
when the ratios $m=M/T$ and $n=N/T$ fixed. As usual with Random Matrix problems \cite{Verdu,Edelman}, 
the singular value spectrum develops sharp edges in the asymptotic limit which are to a large
extent independent of the distribution of the elements of the matrices. 
Any singular value observed to be significantly outside these edges can 
therefore 
be deemed to carry some relevant information. A similar solution has been known for a long time for 
standard correlation matrices, for example the correlations of the $N$ input variables between themselves that define an $N \times N$ symmetric matrix. In this case, the benchmark is known as the Wishart ensemble, and the relevant 
eigenvalue spectrum is given by the Mar\v{c}enko-Pastur distribution \cite{MP,Silverstein,sengupta}. Applications of
this method to financial correlation matrices are relatively recent \cite{PRL} 
but
very active \cite{burda,Cracow}. Comparing the empirical eigenvalues to the correlation matrix to the theoretical upper edge
of the Mar\v{c}enko-Pastur spectrum allows one to extract statistically 
significant factors \cite{PRL} (although some may also be buried below
the band edge, see \cite{burda}). Similar ideas 
are 
starting to be discussed in the econometric community, 
in particular to deal with the problem of identifying the relevant 
factors in large dynamical factor models \cite{Kapetanios}, and 
using them for prediction purposes (see also \cite{Bai}
for a different point of view). 
Here, we extend the Mar\v{c}enko-Pastur
result to general rectangular, non-equal time correlation matrices. We will first present a precise formulation of our 
central result, which we will then illustrate using an economically relevant data set, and finally discuss some possible 
extensions of our work. 

\section{Mathematical formulation of the problem}

We will consider $N$ input factors, denoted as $X_a$, $a=1,...,N$ and $M$ output factors $Y_\alpha$, $\alpha=1,...,M$. 
There is a total of $T$ observations, where both $X_{at}$ and $Y_{\alpha t}$, $t=1,...,T$ are observed. We assume that
all $N+M$ time series are standardized, i.e., both $X$'s and $Y$'s have zero mean and variance unity.  
The $X$ and the $Y$'s may be completely different, or be the same set of observables but observed at different times, 
as for example $N=M$ and $Y_{\alpha t}=X_{at+1}$. From the set of $X$'s and $Y$'s one can form two correlations matrices, 
$C_{X}$ and $C_Y$, defined as:
\be
(C_X)_{ab} = \frac1T \sum_{t=1}^T X_{at} X_{bt} \qquad (C_Y)_{\alpha\beta} = \frac1T \sum_{t=1}^T Y_{\alpha t} Y_{\beta t}. 
\ee
In general, the $X$'s (and the $Y$'s) have no reason to be independent of each other, and the correlation matrices 
$C_X$ and $C_Y$ will contain information on their correlations. As alluded to above, one can diagonalize both these
matrices; provided $T > N,M$ -- which we will assume in the following -- all eigenvalues will, in generic cases, 
be strictly positive. In certain cases, some eigenvalues will however lie close to zero, much below the lower edge 
of the Mar\v{c}enko-Pastur interval, corresponding to redundant variables which may need to be taken care of (see 
below). Disregarding this problem for the moment, we use the corresponding eigenvectors to define a set of uncorrelated, 
unit variance input variables $\hat X$ and output variables $\hat Y$. For example, 
\be\label{defhat}
\hat X_{at} = \frac{1}{\sqrt{T\lambda_a}} \sum_b V_{ab} X_{bt},
\ee
where $\lambda_a$ is the $a^{th}$ eigenvalue of $C_X$ and $V_{ab}$ the components of the corresponding eigenvector.
Now, by construction, $C_{\hat X}=\hat X \hat X^T$ and $C_{\hat Y}=\hat Y \hat Y^T$ are {\it exactly} identity matrices, of dimension, respectively, $N$ and $M$. Using general property 
of diagonalisation, this means that the $T \times T$ matrices $D_{\hat X}=\hat X^T \hat X$ and $D_{\hat Y}=\hat Y^T \hat Y$
have exactly $N$ (resp. $M$) eigenvalues equal to $1$ and $T-N$ (resp. $T-M$) equal to zero. 

Now, consider the $M \times N$ cross-correlation matrix $G$ between the $\hat X$'s and the $\hat Y$'s:
\be 
(G)_{\alpha b} =  \sum_{t=1}^T \hat Y_{\alpha t} \hat X_{bt} 
\equiv \hat Y \hat X^T.
\ee
The singular value decomposition (SVD) of this matrix answers the following question \cite{SVD}: what is the (normalised) 
linear combination of $\hat X$'s on the one hand, and of $\hat Y$'s on the other hand, that have the strongest 
mutual correlation? In other words, what is the best pair of predictor and predicted variables, given the data?
The largest singular value $s_{\max}$ and its corresponding left and right eigenvectors answer precisely this question: the
eigenvectors tell us how to construct these optimal linear combinations, and the associated singular value gives us
the strength of the cross-correlation. One can now restrict both the input and output spaces to the $N-1$ and 
$M-1$ dimensional sub-spaces orthogonal to the two eigenvectors, and repeat the operation. The list of 
singular values $s_a$ gives the prediction power, in decreasing order, of the corresponding linear combinations. 

\section{Singular values from free random matrix theory} 

How to get these singular values? If $M < N$, the trick is to consider the matrix $M \times M$ matrix $GG^T$ (or the $N \times N$ matrix $G^TG$
if $M > N$), which is symmetrical and has $M$ positive eigenvalues, each of which being equal to the 
square of a singular value of $G$ itself. The second observation is that the non-zero eigenvalues of $GG^T=\hat Y \hat X^T
\hat X \hat Y^T$ are the same as those of the $T \times T$ matrix ${\cal D} = D_{\hat X} D_{\hat Y}=\hat X^T \hat X \hat Y^T
\hat Y$, obtained by swapping the position of $\hat Y$ from first to last.
In the benchmark situation where the $\hat X$'s and the $\hat Y$'s are independent from each other, the two matrices
$D_{\hat X}$ and $D_{\hat Y}$ are mutually free \cite{Verdu} and one can use results on the product of free matrices to obtain the
eigenvalue density from that of the two individual matrices, which are known. The general recipe \cite{Silverstein,Verdu} 
is to construct 
first the so-called $\eta-$transform of the eigenvalue density $\rho(u)$ of a given $T \times T$ non negative matrix $A$, 
defined as:
\be
\eta_A(\gamma) = \int {\rm d}u \rho(u) \frac{1}{1+\gamma u} \equiv \frac{1}{T} {\rm Tr}\frac{1}{1 + \gamma A}.
\ee
From the functional inverse of $\eta_A$, one now defines the $\Sigma$-transform of $A$ as:
\be
\Sigma_A(x) \equiv -\frac{1+x}{x} \eta_A^{-1}(1+x).
\ee
Endowed with these definitions, one of the fundamental theorems of Free Matrix Theory \cite{Verdu} states that 
the $\Sigma$-transform
of the product of two free matrices $A$ and $B$ is equal to the product of the two $\Sigma$-transforms. [A similar, somewhat simpler,
theorem exists for {\it sums} of free matrices, in terms of ``R-transforms'', 
see \cite{Verdu}]. Applying this theorem with $A=D_{\hat X}$ and $B=D_{\hat Y}$, one finds:
\be
\eta_A(\gamma)=1-n + \frac{n}{1+\gamma}, \quad n=\frac{N}{T}; \qquad 
\eta_B(\gamma)=1-m + \frac{m}{1+\gamma}, \quad m=\frac{M}{T}.
\ee
From this, one easily obtains:
\be
\Sigma_{\cal D}(x) = \Sigma_A(x) \Sigma_B(x) = \frac{(1+x)^2}{(x+n)(x+m)}.
\ee
Inverting back this relation allows one to derive the $\eta-$transform of ${\cal D}$ as:
\be\label{eta}
\eta_{\cal D}(\gamma)= \frac{1}{2(1+\gamma)} \left[1 - (\mu + \nu) \gamma + \sqrt{(\mu-\nu)^2 \gamma^2 -
2(\mu+\nu+2\mu \nu) \gamma + 1} \right]
\ee
with $\mu=m-1$ and $\nu=n-1$.
The limit $\gamma \to \infty$ of this quantity gives the density of exactly zero eigenvalues, easily found to
be equal to $\max(1-n,1-m)$, meaning, as expected, that the number of non zero eigenvalues of ${\cal D}$ is $\min(N,M)$. 
Depending on the value of $n+m$ 
compared to unity, the pole at $\gamma=-1$ corresponding to eigenvalues exactly
equal to one has a zero weight (for $n+m < 1$) or a non zero weight equal to
$n+m-1$. One can re-write the above result in terms of the more common Stieltjes transform of ${\cal D}$, $S(z) \equiv
\eta(-1/z)/z$, which reads:
\be\label{S}
S_{\cal D}(z)= \frac{1}{2z(z-1)} \left[z + (\mu + \nu) + \sqrt{(\mu-\nu)^2 +
2(\mu+\nu+2\mu \nu) z + z^2} \right].
\ee

\begin{figure}
\begin{center}
\epsfig{file=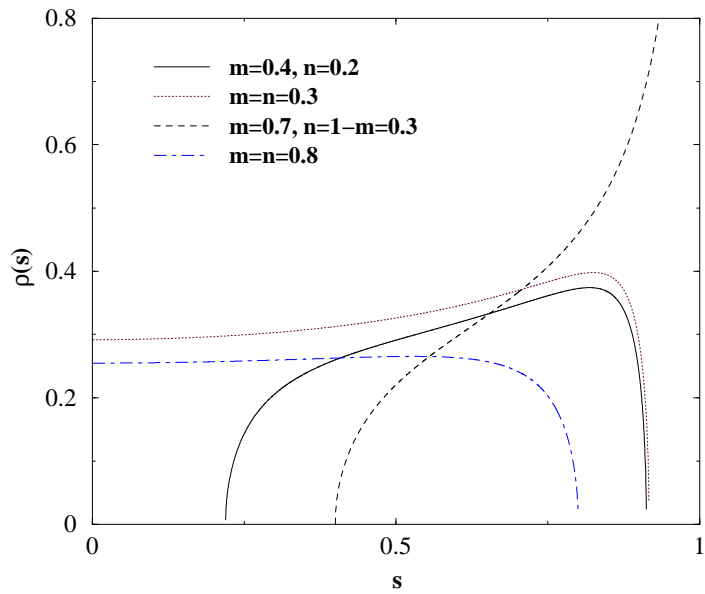,width=8cm,angle=270} 
\end{center}
\caption{\small{
Continuous part of the theoretical random singular value 
spectrum $\rho(s)$ for different values of $n$ and $m$. Note that for
$n=m$ the spectrum extends down to $s=0$, whereas for $n+m \to 1$, 
the spectrum develops a $(1-s)^{-1/2}$ singularity, just before the 
appearance of a $\delta$ peak at $s=1$ of weight $n+m-1$.
}}
\label{Fig0}
\end{figure}

The density of eigenvalues $\rho_{\cal D}(z)$  
is then obtained from the standard relation \cite{Verdu}:
\be
\rho_{\cal D}(z) = \lim_{\epsilon \to 0} \Im \left[\frac{1}{\pi T} {\rm Tr}\frac{1}{z + i \epsilon - {\cal D}}\right]
= \lim_{\epsilon \to 0} \frac{1}{\pi} \Im \left[S_{\cal D}({z + i \epsilon})\right],
\ee
which leads to the rather simple final expression, which is the central result of this paper, for the 
density of singular values $s$ of the original correlation matrix $G\equiv \hat Y \hat X^T$:
\be\label{result}
\rho(s) = \max(1-n,1-m) \delta(s) + \max(m+n-1,0) \delta(s-1) + 
\frac{\Re \sqrt{(s^2-\gamma_-)(\gamma_+-s^2)}}{\pi s(1-s^2)},
\ee
where $\gamma_\pm$ are the two positive roots of the quadratic expression 
under the 
square root in Eq. (\ref{S}) above, which read explicitely:\footnote{One
can check that $\gamma_+ \leq 1$ for all values of $n,m <1$. The upper 
bound is reached only when $n+m=1$, in which case the upper edge of the 
singular value band touches $s=1$.}
\be
\gamma_{\pm} = n+m-2mn \pm 2 \sqrt{mn(1-n)(1-m)}.
\ee
This is our main technical result, illustrated in Fig. 1. 
One can check that in the limit $T \to \infty$ at fixed $N$, $M$, 
all singular values collapse to zero, as they should
since there is no true correlations between $X$ and $Y$; 
the allowed band in the limit $n,m \to 0$ becomes:
\be
s \in \left[|\sqrt{m}-\sqrt{n}|,\sqrt{m}+\sqrt{n}\right].
\ee 
When $n \to m$, the allowed band becomes $s \in [0, 2\sqrt{m(1-m)}]$
(plus a $\delta$ function at $s=1$ when $n+m > 1$),  while when $m=1$, 
the whole band collapses to a $\delta$ function at 
$s=\sqrt{1-n}$. When $n+m \to 1^-$, the inchoate 
$\delta$-peak at $s=1$ is announced as a singularity of $\rho(s)$ 
diverging as $(1-s)^{-1/2}$. Finally, when $m \to 0$ at fixed $n$, 
one finds that the whole band collapses again to a $\delta$ function at 
$s = \sqrt{n}$. This
last result can be checked directly in the case one has one output variable
($M=1$) that one tries to correlate optimally with a set of $N$ independent
times series of length $T$. The result can easily be shown to be a correlation
of  $\sqrt{N/T}$. A plot of the SV density $\rho(s)$ for values of 
$m$ and $n$ which will be used below is shown in Fig 2, 
together with a numerical determination of the SVD spectrum of two independent vector time series $X$ and $Y$, after suitable diagonalisation 
of their empirical correlation matrices to construct their normalised counterparts, $\hat X$ and $\hat Y$. The agreement
with our theoretical prediction is excellent.

\begin{figure}
\begin{center}
\epsfig{file=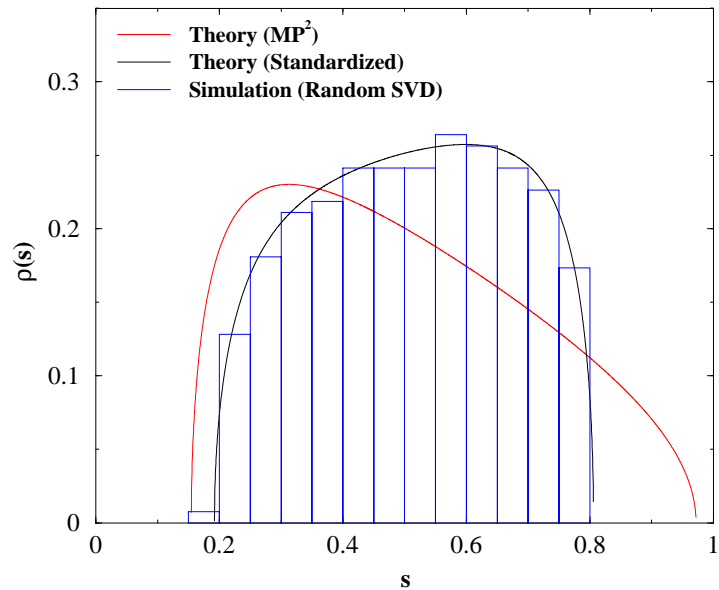,width=8cm,angle=270} 
\end{center}
\caption{\small{
Random Singular Value spectrum $\rho(s)$ for $m=35/265$ and $n=76/265$. We show two possible theoretical 
calculations, corresponding either to bare random vectors $X$ and $Y$, for which the singular value spectrum is 
related to the `square' (in the free convolution sense) of the Mar\v{c}enko-Pastur distribution $MP^2$, or standardized vectors $\hat X$ and $\hat Y$,
obtained after diagonalizing the empirical correlation matrices of $X$ and $Y$. We also show the results of a numerical
simulation of the standardized case with $T=2650$. 
}}
\label{Fig1}
\end{figure}

Note that one could have considered a different benchmark ensemble, where the independent vector time series $X$ and $Y$
are {\it not} diagonalized and transformed into $\hat X$ and $\hat Y$ before SVD. The direct SVD spectrum in that case can also 
be computed as the $\Sigma$-convolution of two Mar\v{c}enko-Pastur distributions with parameters $m$ and $n$, 
respectively (noted $MP^2$ in Fig. 2). The 
result, derived in the Appendix, is noticeably different from the above 
prediction (see Fig. 1). This alternative benchmark ensemble is however
not well suited for our purpose, because it mixes up the possibly non trivial correlation structure of the input variables 
and of the output variables themselves with the issue at stake here, namely the {\it cross}-correlations between 
input and output variables.

\begin{figure}
\begin{center}
\epsfig{file=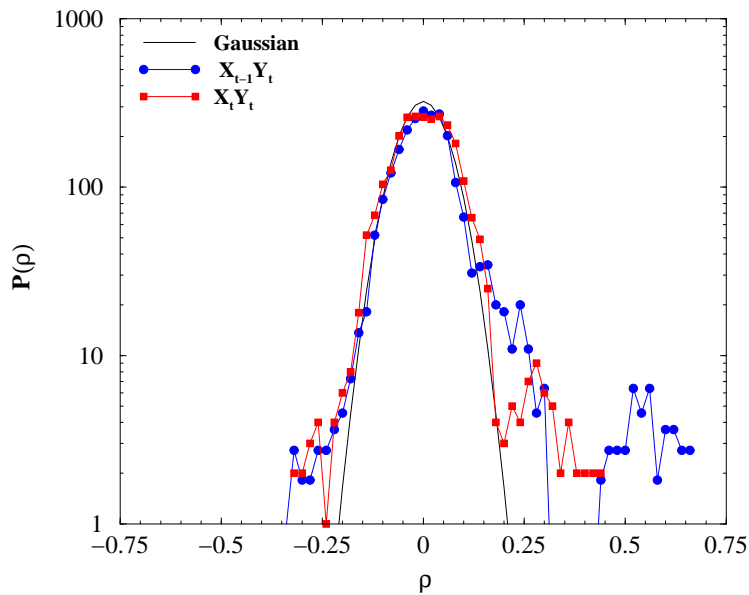,width=8cm,angle=270} 
\end{center}
\caption{\small{
Histogram of the pair correlation coefficient $\rho$ between $X$'s and $Y$'s, both at equal times and 
with one month lag. Note the `island' of correlations around $\approx 0.6$ for one-month lagged correlations, which 
corresponds to correlations between oil prices and energy related CPI's one month later. We also show a Gaussian 
of variance $1/T$, expected in the absence of any correlations.
}}
\label{Fig2}
\end{figure}

\section{Application: inflation vs. economic indicators}

We now turn to the analysis of a concrete example.
We investigate how different groups of US inflation indexes can be explained using combinations 
of indicators belonging to different economic sectors. 
As ``outputs'' $Y$, we use 34 indicators of inflation, the monthly changes of 
the Composite Price Indexes (CPIs),
concerning different sectors of activity including and excluding commodities. These indexes were not 
selected very carefully and some are very redundant, since the point of our study is to show how the 
proposed method is able to select itself the relevant variables. 
As explanatory variables, ``inputs'' $X$, we use 76 different macroeconomic 
indicators from the following categories: industrial production, retail sales, 
new orders and inventory indexes of all available economic activity sectors, 
the most important 
consumer and producer confidence indexes, new payrolls and unemployment difference, interest rates 
(3 month, 2 and 10 years), G7 exchange rates against the Dollar and the WTI oil price itself.
The total number of months in the period June 1983-July 2005 is 265. 
We want to see whether there is any 
significant correlation between changes of the CPIs and of the 
economic indexes, either simultaneous or one month ahead. We also investigated
two-month lag correlations, for which we found very little signal.
 
We first standardized the time series $Y$ and $X$ and 
form the rectangular correlation matrix between these two quantities, containing $34 \times 76$ numbers in the 
interval $[-1,1]$. The distribution of these pair correlations is shown in Fig. 3, both for equal time $\hat Y_t 
\hat X_t'$ and for one-month 
lagged $\hat Y_t \hat X_{t-1}^T$ correlations, and compared to a Gaussian distribution of variance $T^{-1}$. 
We see 
that the empirical distributions are significantly broader; in particular an `island' of correlations around 
$\approx 0.6$ appears for the one-month lagged correlations. These correspond to correlations between oil prices and energy related CPIs one month later.
The question is whether there are other predictable modes in the system, in particular, are the correlations in the left and right
flanks of the central peak meaningful or not? This question is {\it a priori} non trivial because the kurtosis of 
some of the variables is quite high, which is expected to `fatten' the 
distribution of $\rho$ compared to the Gaussian. 
Within the period of about thirty years covered by our time series, three 
major rare events happened: the Gulf War (1991-92), the Asian crisis (1998), and the Twin Towers Attack (2001). 
The kurtosis of the CPIs is the trace of the corresponding outliers, such as the food price index and its `negative', 
the production price index excluding food, which are strongly sensitive to war events. 
Among economic indicators, the most responsive series to these events appear to be the inventory-sales ratio, 
the manufacturing new orders and the motor and motor parts industrial production indexes.

In order to answer precisely the above question, we first turn to the analysis of the empirical self-correlation 
matrices $C_X$ and $C_Y$, which we diagonalize and represent the eigenvalues compared to the 
corresponding Mar\v{c}enko-Pastur 
distributions in Fig. 4, expected if the variables were independent (see the Appendix for more
details). Since the both the input and output variables 
are in fact rather strongly correlated at equal times, it is not surprising to find that some large eigenvalues $\lambda$ 
emerge from the Mar\v{c}enko-Pastur noise band: for $C_X$, the largest eigenvalue is $\approx 15$, to be compared to 
the theoretical upper edge of the Mar\v{c}enko-Pastur distribution $2.358$, whereas for $C_Y$ the largest eigenvalue is 
$\approx 6.2$ to be compared with $1.858$. But the most important point for our purpose is the rather large number of 
very small eigenvectors, much below the lower edge of the Mar\v{c}enko-Pastur distribution ($\lambda_{\min}=0.215$ for
$C_X$, see Fig. 4). These 
correspond to linear combinations of redundant (strongly correlated) indicators. 
Since the definition of $\hat X$ and $\hat Y$ include 
a factor $1/\sqrt{\lambda}$ (see Eq. (\ref{defhat})), the eigenvectors corresponding to these small eigenvalues have 
an artificially enhanced weight. One expects this to induce some extra noise in the system, as will indeed be clear below.
Having constructed the set of strictly uncorrelated, unit variance input $\hat X$  and output $\hat Y$ variables, we 
determine the singular value spectrum of $G = \hat Y \hat X^T$.  If we keep all variables, this spectrum is in fact
indistinguishable from pure noise when $\hat X$ precedes $\hat Y$ by one month, and only one eigenvalue emerges 
($s_{\max} \approx 0.87$ instead of the theoretical value $0.806$) when $\hat X$ and $\hat Y$ are simultaneous. 

If we now remove redundant, noisy factors that correspond to, say,  $\lambda \leq \lambda_{\min}/2 \approx 0.1$ 
both in $C_X$ and $C_Y$, we reduce the number of factors to $50$ for $\hat X$ and $16$ for $\hat Y$ \footnote{The results we find are however weakly dependent
on the choice of this lower cut-off, provided very small $\lambda$'s are 
removed.}. The cumulative
singular value spectrum of this cleaned problem is shown in Fig. 5 and compared again to the corresponding random
benchmark. In this case, both for the simultaneous and lagged cases, the top singular values $s_{\max} \approx 0.73$ 
(resp. $s_{\max} \approx 0.81$) are very clearly above the theoretical upper edge $s_{ue} \approx 0.642$, 
indicating the presence of 
some true correlations. The top singular values $s_{\max}$ rapidly sticks onto the theoretical edge as the lag 
increases. For the one-month lagged case, there might be a second 
meaningful singular value at 
$s=0.66$. The structure of the corresponding eigenvectors allows one to construct a linear combination of 
economic indicators explaining a linear combinations of CPIs series. The combination of economic indicators 
corresponding to the top singular value evidences the main economic factors 
affecting inflation indicators: oil prices obviously correlated to energy production increases 
and electricity production decreases that explain the CPIs indexes including oil and energy.
The second factor includes the next important elements of the economy: 
employment (new payrolls) affects directly the ``core'' indexes and the CPI indexes excluding oil. 
New economy production (high tech, media \& communication) is actually a proxy for productivity 
increases, 
and therefore exhibits a negative correlation with the same core indexes. 
We have also computed the inverse participation ratio of all left and right 
eigenvectors with similar conclusions \cite{PRL}: all eigenvectors have a participation
ratio close to the informationless Porter-Thomas result, 
except those corresponding to singular values above the upper edge. 

\begin{figure}
\begin{center}
\epsfig{file=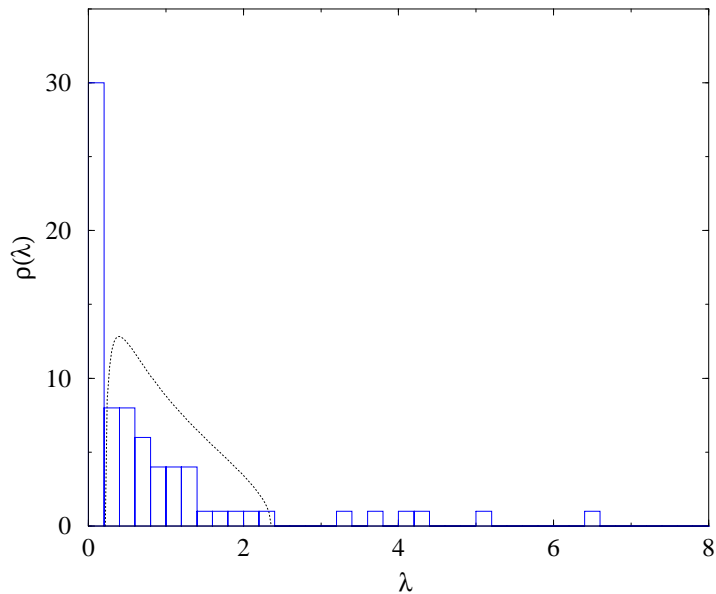,width=8cm,angle=270} 
\end{center}
\caption{\small{Eigenvalue spectrum of the $N \times N$ correlation matrix of the input variables $C_X$, compared to 
the Mar\v{c}enko-Pastur distribution with parameter $n=76/265$. Clearly, 
the fit is very bad, meaning that the input variables are strongly 
correlated; the top eigenvalues $\lambda_{\max} \approx 15$ is in fact 
not shown. Note the large number of very small 
eigenvectors corresponding to combinations
of strongly correlated indicators, that are pure noise but have a 
small volatility.  
}}
\label{Fig3}
\end{figure}

\begin{figure}
\begin{center}
\epsfig{file=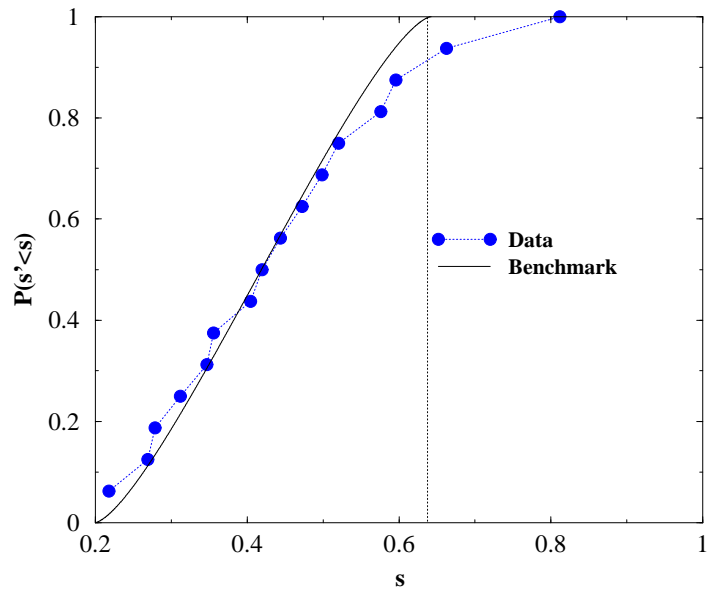,width=8cm,angle=270} 
\end{center}
\caption{\small{Cumulative singular value distribution for the ``cleaned'' problem, i.e. removing the factors with
very small volatilities, leaving $50$ factors in $\hat X$ and $16$ in $\hat Y$. The correlations we consider are
lagged and correspond to $\hat Y_t \hat X_{t-1}^T$. The filled circles 
correspond to the $16$ empirical singular values, 
and the plain line is the theoretical prediction in the purely random case with $n=50/265$ and $m=16/265$. Note that the 
top singular value  $s_{\max} \approx 0.81$ clearly stands out of the noise band, the edge of which is at $s_{ue}=0.642$. 
Finite $T$ corrections are expected to smooth the edge over a region of size $T^{-2/3} \approx 0.025$ for $T = 265$.
}}
\label{Fig4}
\end{figure}

Since $Y_{t-1}$ may also contain some information to predict $Y_t$, 
one could also study, in the spirit of 
general Vector Autoregressive Models \cite{Gew,Forni,SW05}, 
the case where we consider the full vector of observables
$Z$ of size $111$, obtained by merging together $X$ and $Y$. 
We again define the normalised vector $\hat Z$, remove all redundant eigenvalues of $\hat Z \hat Z'$ smaller than $0.1$, 
and compute the singular value spectrum of $\hat Z_t \hat Z_{t-1}^T$. The size of this cleaned matrix is $62 \times 62$, 
and the upper edge of the random singular value spectrum is $s_{ue} \approx 0.84$. We now find that the top 
singular value is at $s_{\max} \approx 0.97$, and that  $\sim 8$ factors have singular values above the upper edge of 
the random spectrum. The top singular value corresponds to sales and inventory/sales ratio, followed by CPIs that tend 
to be correlated over time. Further results are less intuitively simple. This analysis can of course be generalized to 
larger lags, by studying $\hat Z_t \hat Z_{t-n}^T$. We find that even for $n=4$, there are still three singular values 
above the upper edge. The SVD results are therefore of great help to rank the importance 
of autocorrelations of degree $n$ in the system; we will explore this point further in a future
publication.

\section{Conclusions and extensions}
 
The conclusions of this illustrative empirical study are twofold: (i) in general, both input and output variables have a
non trivial correlation structure, with many redundant factors which add a significant amount of noise in the
problem. Therefore, in a first step, some data cleaning must be performed by eliminating these redundant 
variables; (ii) the singular value spectrum, compared to its purely random counterpart, allows one to answer precisely 
the question of the number and relevance of independent predictable factors in the problem under study. In the
case considered, we have seen that although the number of pairs of apparently correlated factors is large (see Fig. 3), 
only a few modes can in fact be considered as containing useful information, in the sense that their singular value
exceeds our analytical upper edge given in Eq. (\ref{result}). When studying the full problem where all variables are treated 
together, we find that the effective dimensionality of the problem drops from $111$ to eight or so independent, 
predictable factors. This compares quite well with the number seven quoted by Stock and Watson within their dynamical
factor analysis of a similar data set \cite{SW05}. A more thorough comparison of our results with those of the 
econometrics literature will be presented elsewhere.

What we mean by `exceed the upper edge' should of course be specified more 
accurately, beyond the eye-balling procedure that we implicitly rely on. In order to have a more precise criterion, one
should study the statistics of the top eigenvalue of ${\cal D}$, which is, in analogy with the known results for 
the Wishart ensemble, most probably given by a Tracy-Widom
distribution, at least for Gaussian random variables (see \cite{BenArous,ElKaroui} for recent progress and references).
For finite $T$, we expect the top eigenvalue of ${\cal D}$ to ooze away from the theoretical edge by a quantity 
of order $T^{-2/3} \approx 0.025$ for $T = 265$. Therefore, the difference between $s_{\max} \approx 0.81$ and the
theoretical edge $s_{ue}=0.642$ reported in Fig. 5 can safely be considered as significant when all variables are Gaussian.
However, although the density of singular values is to a large degree independent of the distribution
of the matrix entries, one should expect that the fuzzy region around the theoretical edge 
expands significantly if the input and output variables have fat tails. 
In particular, the Tracy-Widom distribution is expected to breakdown in some way that
would be very interesting to characterize precisely. We leave this problem to future investigations. 

In conclusion, we have presented a general method to extract statistically meaningful correlations between an
arbitrary collection of input and output variables of which only a finite time sample is available. Our central 
result is derived from the theory of free random matrices, and gives an explicit expression for the interval where
singular values are expected {\it in the absence of any true correlations} between the variables under study.
Our result can be seen as the natural generalization of the Mar\v{c}enko-Pastur distribution for the case of 
rectangular correlation matrices. The potential applications of this method are quite numerous and we hope that 
our results will prove useful in different fields where multivariate correlations are relevant.

\vskip 1cm 

Acknowledgments: We wish to thank G\'erard Ben Arous, Florent Benaych-Georges and Jack Silverstein for most useful 
discussions on Random Matrix Theory. 

\section*{Appendix: the $MP^2$ case}

As indicated in the main text, one could have chosen as a benchmark the case where all (standardized) 
variables $X$ and $Y$ are uncorrelated, meaning that the ensemble average  $E(C_X)=E(XX^T)$ and $E(C_Y)=E(YY^T)$
are equal to the unit matrix, whereas the ensemble average cross-correlation $E(G)=E(YX^T)$ is 
identically zero. However, for a given finite size sample, the eigenvalues of $C_X$ and $C_Y$ 
will differ from unit, and the singular values of $G$ will not be zero. The statistics of the 
eigenvalues of $C_X$ and $C_Y$ is well known to be given by the Mar\v{c}enko-Pastur distribution
with parameters $n$ and $m$ respectively, which reads, for $\beta=n,m < 1$:
\be
\rho_{MP}(\lambda)= \frac{1}{2 \pi \beta \lambda} \Re \sqrt{(\lambda-\lambda_{\min})(\lambda_{\max}
-\lambda)},
\ee 
with 
\be
\lambda_{\min}=(1 - \sqrt{\beta})^2 \qquad \lambda_{\max}=(1 + \sqrt{\beta})^2.
\ee
The $\Sigma$-transform of this density takes a particularly simple form:
\be
\Sigma(x)=\frac{1}{1+\beta x}.
\ee
Now, as explained in the main text, the singular values of $G$ are obtained as the square-root of 
the eigenvalues of $D=X^TXY^TY$. Since $X^TX$ and $Y^TY$ are mutually free, one can again use the 
multiplication rule of $\Sigma$-transforms, after having noted that the $\Sigma$-transform of the
$T \times T$ matrices $X^TX$ and $Y^TY$ are now given by:
\be
\Sigma(x)=\frac{1}{\beta+ x}.
\ee
One therefore finds that the $\eta$ transform of $D$ is obtained by solving the following 
cubic equation for $x$:
\be
\eta^{-1}(1+x)= - \frac{1+x}{x(n+x)(m+x)},
\ee
which can be done explicitely, leading to the following (lengthy) result.
Denote $z=s^2$, one should first compute the following two functions:
\be 
f_1(z)=1 + m^2 +  n^2 -mn - m -n + 3z
\ee
and
\be
f_2(z)=2- 3m(1-m) - 3n(1-n) -3mn(n+m-4) + 2(m^3+n^3) + 9z(1 + m + n).
\ee
Then, form:
\be
\Delta=-4 f_1(z)^3 + f_2(z)^2.
\ee
If $\Delta > 0$, one introduces a second auxiliary variable
$\Gamma$:
\be
\Gamma=f_2(z) - \sqrt{\Delta},
\ee
to compute $\rho_2(z)$:
\be
\pi \rho_2(z)=-\frac{\Gamma^{1/3}}{2^{4/3}3^{1/2}z} +
\frac{f_1(z)}{2^{2/3}3^{1/2}\Gamma^{1/3}z}.
\ee
Finally, the density $\rho(s)$ is given by:
\be
\rho(s)=2s \rho_2(s^2).
\ee

\end{document}